\documentclass{my-ws-procs9x6}

\begin{document}

\title{\null\vspace*{-2cm}\null
Search for SM and MSSM Higgs Bosons at LEP\footnote{
\uppercase{T}alk presented at {\it \uppercase{SUSY} 2003:
\uppercase{S}upersymmetry in the \uppercase{D}esert}\/, 
held at the \uppercase{U}niversity of \uppercase{A}rizona,
\uppercase{T}ucson, \uppercase{AZ}, \uppercase{J}une 5-10, 2003.
\uppercase{T}o appear in the \uppercase{P}roceedings.}
}

\author{
GABRIELLA P\'ASZTOR\footnote{\uppercase{O}n leave from 
\uppercase{KFKI RMKI, B}udapest, \uppercase{H}ungary.}}

\address{Physics Department,
University of California,
Riverside, CA 92521, USA \\
Mailing address:
CERN, Gen\`eve 23, CH-1211, Switzerland \\
E-mail: 
Gabriella.Pasztor@cern.ch}


\maketitle

\null
\vspace*{-7.5cm}

\begin{flushright}
OPAL Conference Report CR520 \\
11 March, 2004
\end{flushright}

\vspace*{6cm}

\abstracts{Latest results from the LEP Collaboration on searches for neutral 
Higgs bosons predicted by the Standard Model and its minimal 
supersymmetric extension, the MSSM, are summarized.}

\section{Introduction}

The four LEP experiments (ALEPH, DELPHI, L3 and OPAL) collected around 2.5
fb$^{-1}$ data in total at energies $\sqrt{s} \geq 189$ GeV of which 536
pb$^{-1}$ was registered at $\sqrt{s}= 206 - 209$ GeV in 2000. 
The data are used to search for Higgs bosons in a variety of models.

In the Standard Model (SM) the electroweak (EW) symmetry is broken via the
Higgs mechanism generating the masses of elementary particles. This implies the
existence of a single neutral scalar particle, the Higgs boson. While the SM is
successful in describing the observed phenomena there are several theoretical 
arguments requiring its extension. The minimal supersymmetric (SUSY) extension
of the SM (MSSM) introduces two Higgs field doublets leading to five Higgs
bosons: three neutral and two charged.

\section{Standard Model}

At LEP energies, the SM Higgs boson is produced via the Higgs-strahlung process
e$^+$e$^-$ $\rightarrow$ HZ. Vector-boson fusion processes e$^+$e$^-$
$\rightarrow$ He$^+$e$^-$ and H$\nu_{\mathrm e}\bar\nu_{\mathrm e}$ play also
some role close to the kinematic limit. The Higgs boson is expected to decay
predominantly into b\=b with some contribution to $\tau^+\tau^-$, gg,
W$^*$W$^*$ and c\=c. For a Higgs mass of 115 GeV, more than half of the Higgs
events are expected in the four-jet channel (HZ $\rightarrow$ b\=bq\=q). 

\begin{figure}
\centerline{\epsfysize=1.5in\epsfbox{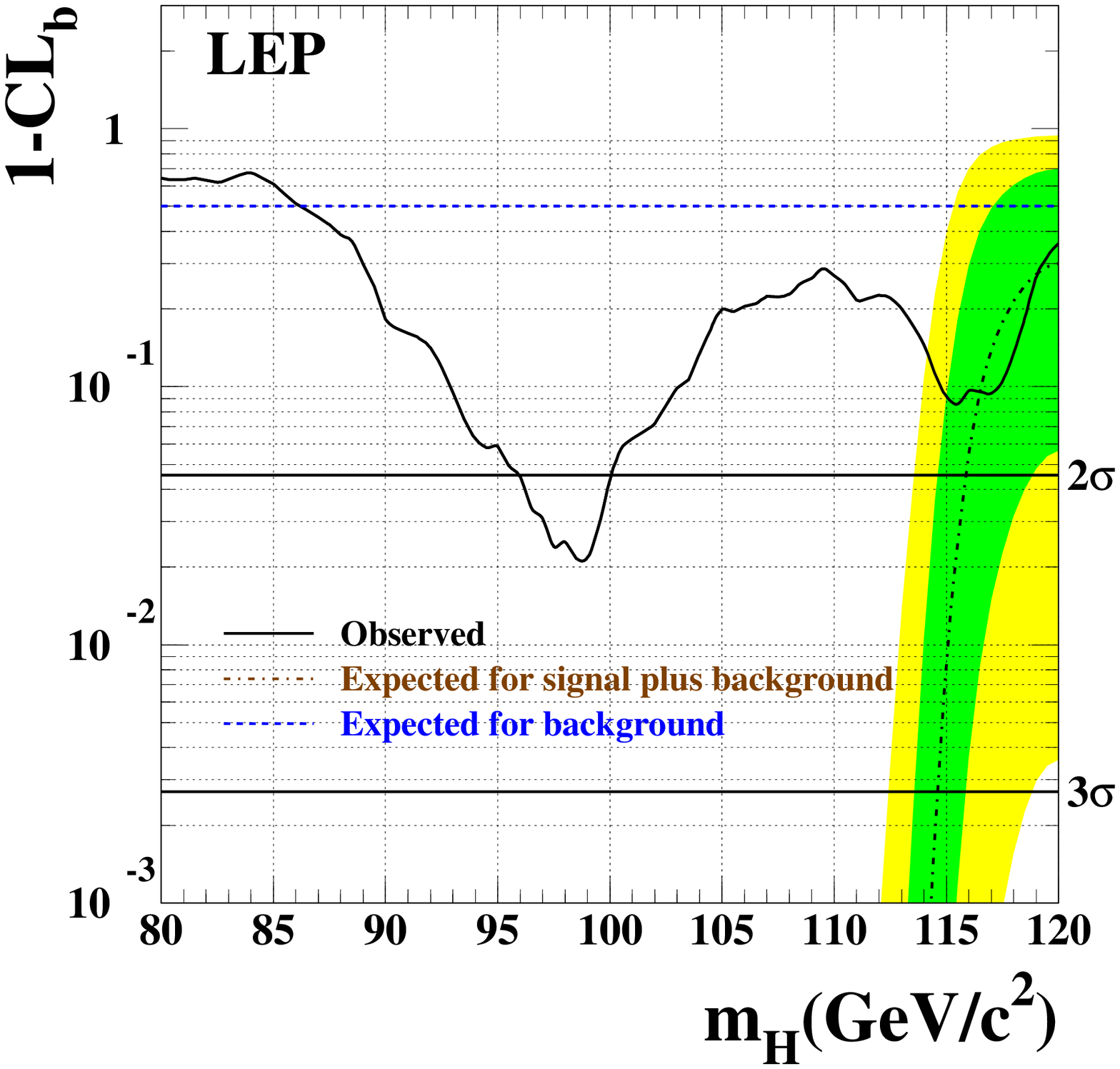}
\epsfysize=1.5in\epsfbox{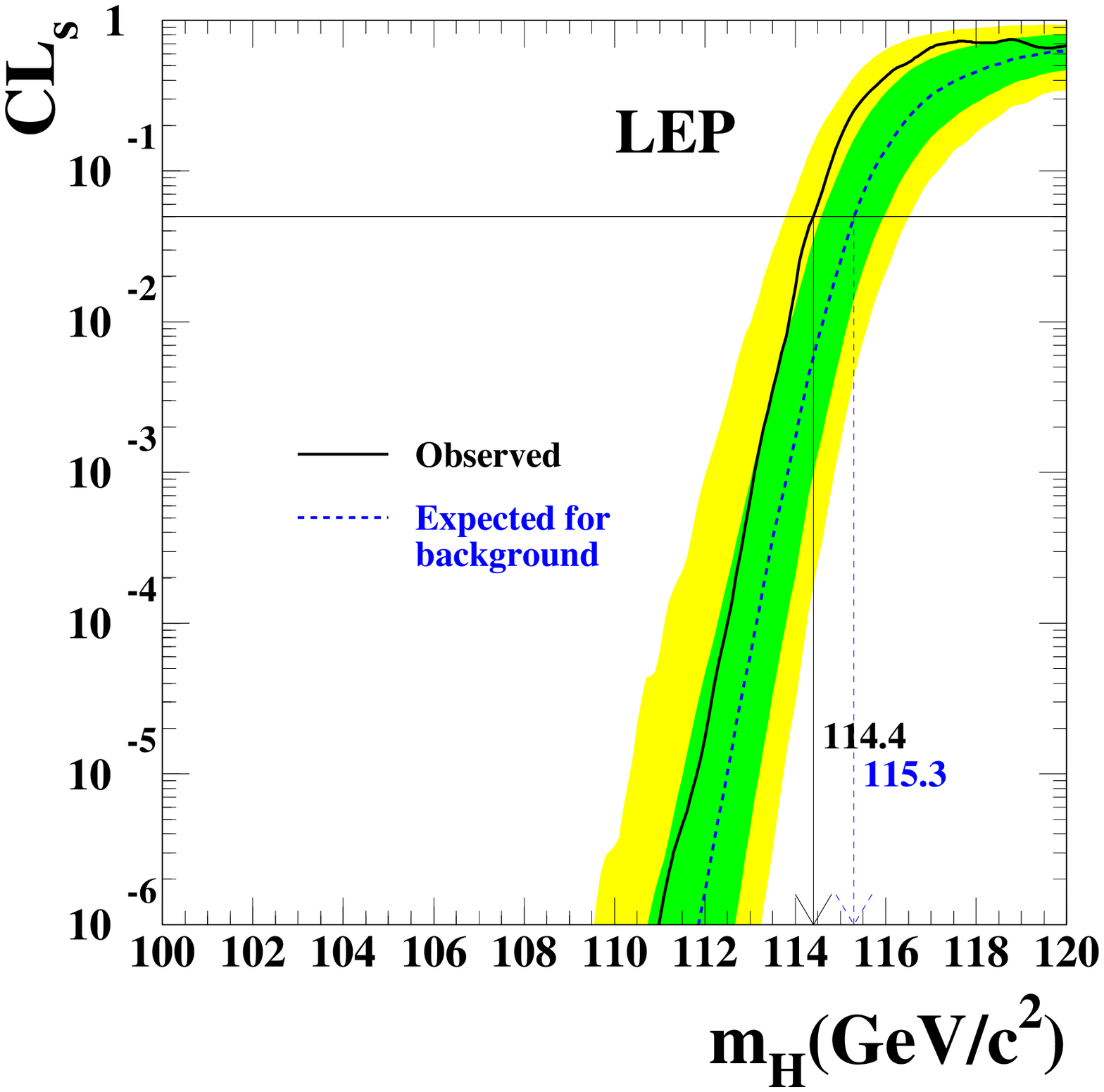}
\epsfysize=1.5in\epsfbox{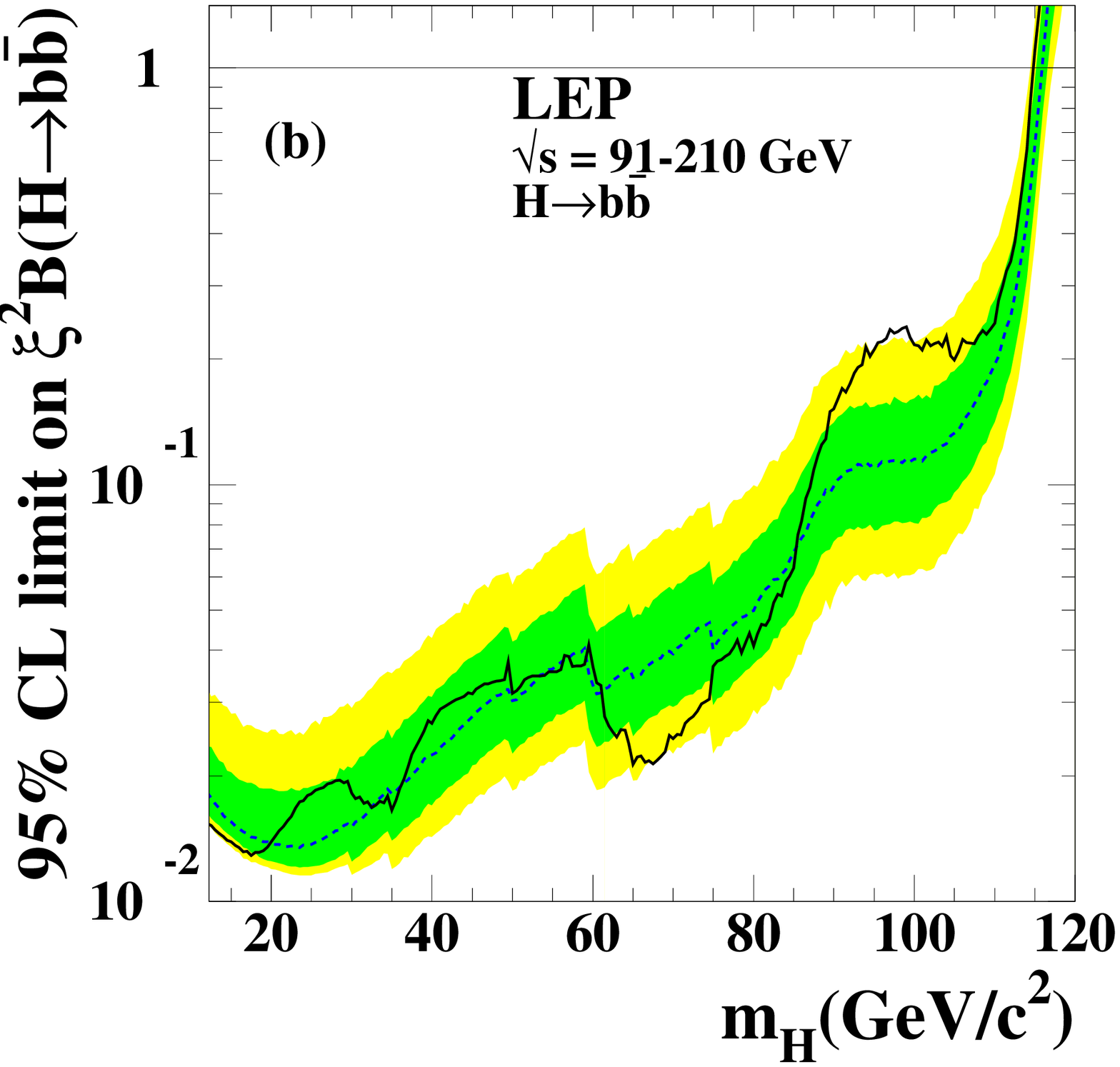}\vspace*{-3mm}
}
\caption{Search for the SM Higgs boson. 
(a) The background confidence 1-CL$_{\mathrm b}$. 
(b) The ratio CL$_{\mathrm s}$=CL$_{\mathrm s+b}$/CL$_{\mathrm b}$ for the
signal-plus-background hypothesis.
(c) The 95\% CL upper bound on the coupling ratio squared 
$\xi^2 = (g_{\mathrm HZZ} / g_{\mathrm HZZ}^{\mathrm SM})^2$ assuming that the
Higgs boson decays exclusively into b\=b.
}
\label{fig:SM}
\end{figure}

During the last year of LEP operation the ALEPH collaboration observed an
excess of four-jet events compatible with SM Higgs boson production with a
Higgs mass of 115 GeV. For this mass hypothesis, the ALEPH data give a
background confidence 1$-$CL$_{\mathrm b}$ = 3.3$\times$10$^{-3}$,
corresponding to an approximately 3$\sigma$ excess, and a
signal-plus-background confidence CL$_{\mathrm s+b} = 0.87$. When combined with
the results of the other LEP experiments\cite{SM} the significance of the
excess decreases below 2$\sigma$, and the confidences are 1$-$CL$_{\mathrm b}$
= 0.09 (see Figure~\ref{fig:SM} (a)) and CL$_{\mathrm s+b}$ = 0.15. A lower
bound of 114.4 GeV is set on the mass of the SM Higgs  boson at the 95\% CL
with an expected limit of 115.3 GeV as shown on Figure~\ref{fig:SM} (b). The
data are also used to derive limits on the HZZ coupling for various assumptions
concerning the Higgs boson decay properties (see Figure~\ref{fig:SM} (c) for
the assumption of 100\% H $\rightarrow$ b\=b decay). 
\vspace*{-5mm}

\vspace*{-5mm}

\section{MSSM}

In the MSSM, the Higgs potential is assumed to be invariant under CP
transformation at tree level. It is possible, however, to break CP symmetry in
the Higgs sector by radiative corrections. Such a scenario provides a possible
solution to the cosmic baryon asymmetry.

Both CP conserving (CPC) and CP violating (CPV) scenarios are studied at LEP. 
In the CPC case, the three neutral Higgs bosons are CP eigenstates: h and H are
CP even, A is CP odd. They are mainly produced in the Higgs-strahlung
processes  e$^+$e$^-$ $\rightarrow$ hZ and HZ and the pair-production processes
e$^+$e$^-$ $\rightarrow$ hA and HA. In the CPV case, however, the three neutral
Higgs bosons, H$_i$, are mixtures of CP-even and CP-odd Higgs fields and the
e$^+$e$^-$ $\rightarrow$ H$_i$Z and H$_i$H$_j$ ($i,j=1,2,3, i \ne j$) processes
may all occur. The decay properties of the Higgs bosons, while quantitatively
different in the two scenarios, maintain a certain similarity: the largest
branching ratios are those to b\=b and $\tau^+\tau^-$ and cascade decays (h
$\rightarrow$ AA and H$_2$ $\rightarrow$ H$_1$H$_1$) occur and can even be
dominant when kinematically allowed.

A large number of search channels are used in the MSSM Higgs hunt: SM Higgs
searches are reinterpreted, searches for hA pair-production,  bbh and bbA
Yukawa productions, flavour independent hZ and hA, decay mode independent hZ
searches are developed. h $\rightarrow$ AA and A $\rightarrow$ hZ decays are
considered in cascade decays leading to six-fermion final states. The search
for invisible decay of Higgs bosons is also useful to explore specific areas of
the MSSM parameter space.
In general, searches designed to detect CPC Higgs production can be
reinterpreted in the CPV scenario. However, in some parts of the CPV parameter
space modified or newly developed searches are also necessary to cover new 
final state topologies.  

The results of the different Higgs searches are interpreted in the framework of
a constrained MSSM with seven parameters. At tree level two parameters are
sufficient to describe the Higgs sector, they are chosen to be the ratio of the
vacuum expectation values (tan$\beta$) and a Higgs mass ($m_{\mathrm A}$ in CPC
and $m_{\mathrm H^\pm}$ in CPV scenarios). Additional parameters appear after
the radiative corrections: the soft SUSY breaking parameter in the sfermion
sector at the EW scale ($m_{\mathrm SUSY}$), the SU(2) gaugino mass parameter
($M_2$), the common trilinear Higgs-squark coupling parameter ($A$), the gluino
mass ($m_{\mathrm {\tilde g}}$) and the SUSY Higgs mass parameter ($\mu$).

Instead of varying all the above parameters, only a certain number of
representative benchmark sets are considered where the tree level parameters
are scanned while all other parameters are fixed. On top of the traditional
three LEP benchmark scenarios (large-$\mu$, no-mixing and m$_{\mathrm h}$-max)
several new scenarios, motivated by limits from b $\rightarrow$ s$\gamma$ and
$(g-2)_\mu$ measurements or by future searches to be conducted at the LHC, are
studied. The recently proposed CPV scenario, called CPX, and its several
derivates are also probed.

\begin{figure}
\centerline{\epsfysize=1.4in\epsfbox{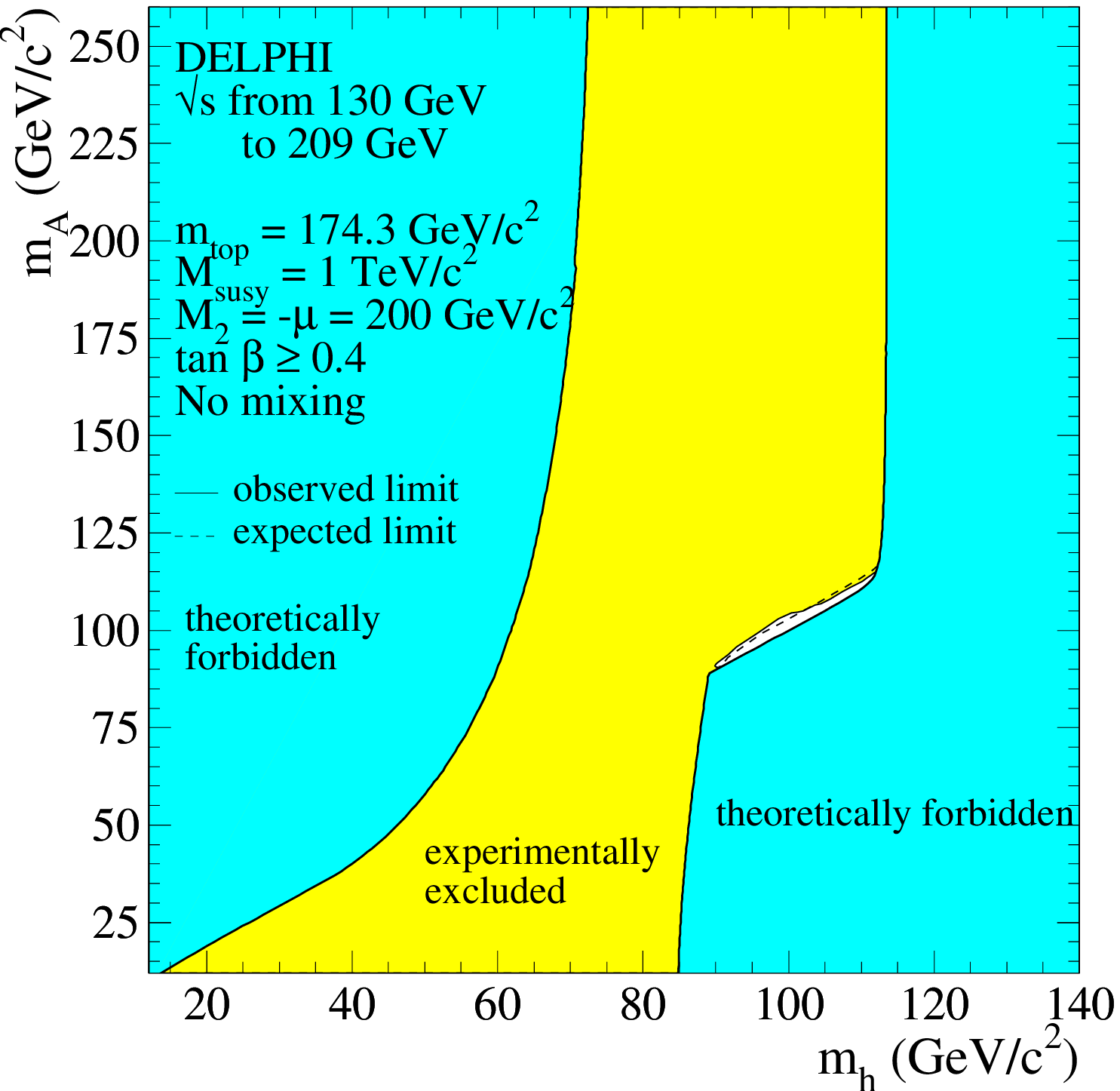}\hspace*{4mm}
\epsfxsize=1.5in\epsfysize=1.4in\epsfbox{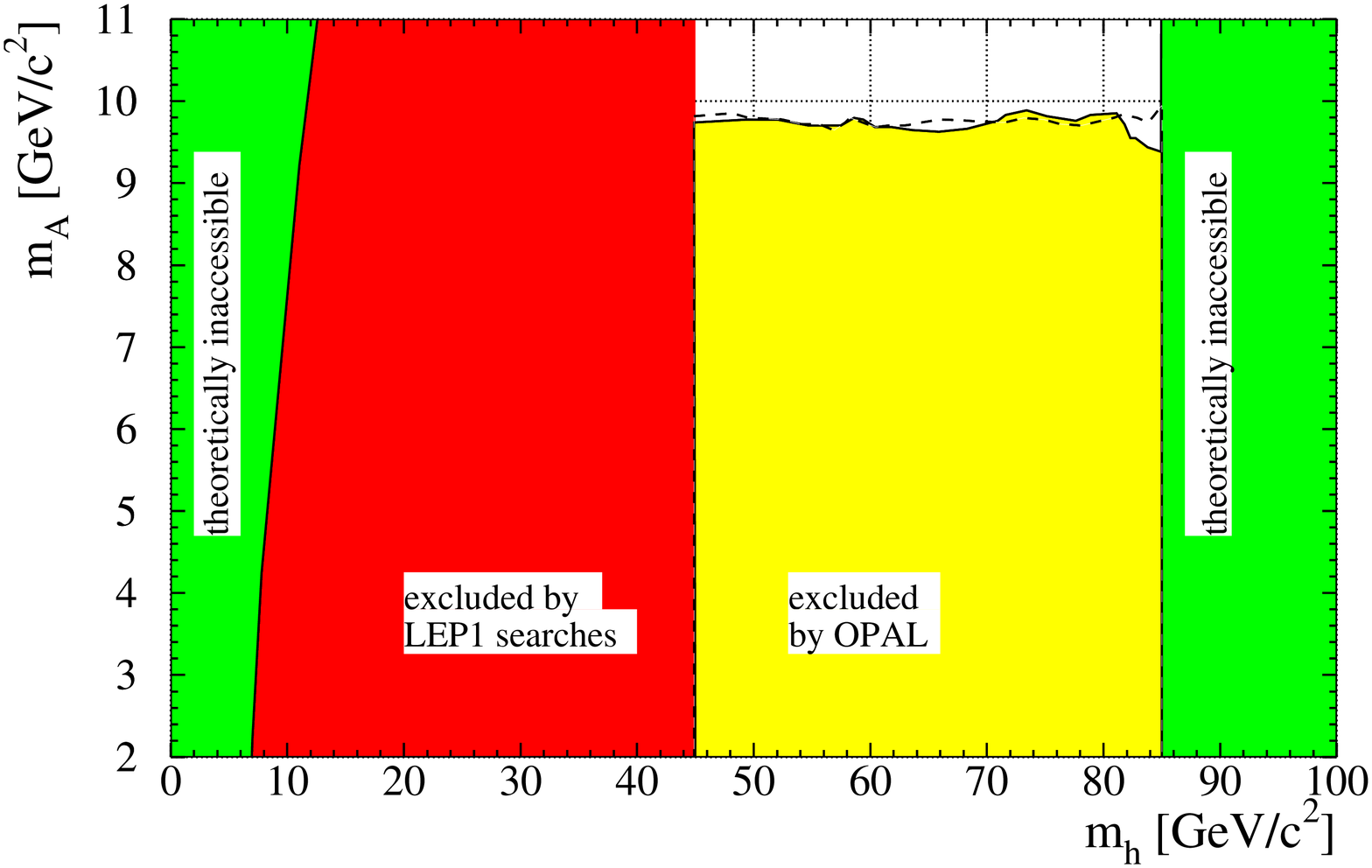}\vspace*{-2mm}
}
\caption{Search for the MSSM Higgs boson, no-mixing scenario. 
Exclusion in the m$_{\mathrm h}$ -- m$_{\mathrm A}$ plane (a) by the DELPHI
collaboration using traditional hZ and hA searches and (b) by the OPAL 
collaboration using a dedicated search for hZ $\rightarrow$ AAZ for
low m$_{\mathrm A}$.
}
\label{fig:MSSM-nomixing}
\end{figure}

The large-$\mu$ scenario, designed to be kinematically always accessible but to
have the h $\rightarrow$ b\=b decay suppressed, is entirely excluded by the
preliminary LEP combination\cite{MSSM} thanks to flavour independent searches.
The no-mixing scenario where the parameters are arranged to have no mixing
between the left- and right-handed stop fields, is strongly constrained even by
a single experiment\cite{MSSM-delphi,MSSM-lowma} as shown in
Figure~\ref{fig:MSSM-nomixing}. In the m$_{\mathrm h}$-max scenario which gives
the maximal value of $m_{\mathrm h}$ for given tan$\beta$ and $m_{\mathrm A}$,
the lower limit on the Higgs boson masses are $m_{\mathrm h} > 91.0$ GeV and
$m_{\mathrm A} > 91.9$ GeV and the range $0.5 <$ tan$\beta$ $<$ 2.4 is
excluded\cite{MSSM}. Studies from the OPAL experiment show that the newly
proposed CPC benchmark scans do not present new difficulties at LEP: the
derived lower limits on the Higgs boson masses\cite{MSSM-opal} in the new
benchmark scenarios vary between 79.0 and 84.5 GeV for h and 84.0 and 90.0 GeV
for A. 
\vspace*{-3mm}

\vspace*{-4mm}

\begin{figure}
\centerline{\epsfysize=1.5in\epsfbox{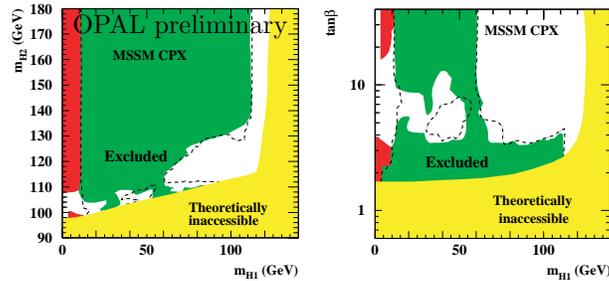}\vspace*{-38mm}
}
\hspace*{25mm}OPAL preliminary\vspace*{30mm}
\caption{Search for the MSSM Higgs boson, CPX scenario. 
Exclusion in the (a) m$_{\mathrm H_1}$ -- m$_{\mathrm H_2}$ and (b)
m$_{\mathrm H_1}$ -- tan$\beta$ planes.
}
\label{fig:MSSM-CPX}
\end{figure}

The CPX scenario with maximal CP violation in the Higgs sector shows a
decoupling of H$_1$  from the Z in the range 4 $<$ tan$\beta$ $<$ 10. H$_2$
couples to the Z and heavier than around 100 GeV. Where kinematically open
H$_2$ $\rightarrow$ H$_1$H$_1$ is dominant. The excluded areas\cite{MSSM-opal}
are shown in Figure~\ref{fig:MSSM-CPX}.  
\vspace*{-3mm}

\vspace*{-6mm}


\end{document}